\documentclass[]{interact}

\usepackage{epstopdf}% To incorporate .eps illustrations using PDFLaTeX, etc.
\usepackage{subfigure}% Support for small, `sub' figures and tables

\usepackage[numbers,sort&compress,merge]{natbib}% Citation support using natbib.sty
\bibpunct[, ]{(}{)}{,}{n}{,}{,}% Citation support using natbib.sty
\renewcommand\bibfont{\fontsize{10}{12}\selectfont}% Bibliography support using natbib.sty
\renewcommand\citenumfont[1]{\textit{#1}}% Citation numbers in italic font using natbib.sty
\renewcommand\bibnumfmt[1]{(#1)}% Parentheses enclose ref. numbers in list using natbib.sty

\theoremstyle{plain}% Theorem-like structures

\theoremstyle{definition}

\theoremstyle{remark}

\usepackage{braket}
\newcommand{\identity}{\mbox{1}\hspace{-0.25em}\mbox{l}}

\usepackage[dvipdfmx, colorlinks=true]{hyperref}
\usepackage[]{comment}

\begin{document}

%\articletype{RESEARCH ARTICLE}

\title{Electromagnetism at finite temperature: a density operator approach}

\author{
\name{D. Oue\textsuperscript{a $\ast$}\thanks{$^\ast$ CONTACT D. Oue. Email: daigo.oue@gmail.com}}
\affil{\textsuperscript{a}The Blackett Laboratory, Department of Physics, Imperial College London, Prince Consort Road, Kensington, London SW7 2AZ, United Kingdom}
}

\maketitle

\begin{abstract}
  In order to analyse classical electromagnetism in a medium at finite temperature we introduce `an optical density operator', and reformulate Maxwell's equations with the operator, 
  starting from the Dirac-equation-like formulation of electromagnetism.
  We find the thermal state of electromagnetic field in the medium from the `optical Dirac Hamiltonian',
  which is the effective Hamiltonian in the Dirac-like formulation.
  In the thermal state,
  the two transverse modes (left-handed and right-handed circular polarisation) of electromagnetic fields exist at the same ratio.
  We also analyse the asymptotics of the thermal state.
  At the low temperature limit,
  there is correlation between the electric field and the magnetic field.
  This means that there exists an electromagnetic wave at the thermal equilibrium,
  and this recovers Maxwell's classical electromagnetism.
  In contrast, the correlation vanishes at the high temperature limit.
  This means that electromagnetic waves are unsustainable and only independent electric fields and magnetic fields exist at the high temperature limit.
\end{abstract}

\begin{keywords}
Classical electromagnetism; optical Dirac equation; thermal state; decoherence
\end{keywords}

\section{Introduction}
\label{sec:introduction}
\par
There have been many attempts to seek simpler formulation of Maxwel's equations,
for twenty equations were used in original works by Maxwell in 1865 \cite{maxwell1865a, rautio2014long}.
One of the most popular forms of Maxwell's equations today is 
\begin{align}
  \nabla \cdot \bm{D} &= 0, \label{eq:Gauss}\\
  \nabla \cdot \bm{B} &= 0, \label{eq:monopole}\\
  \nabla \times \bm{E} &= -\frac{1}{c} \frac{\partial \bm{B}}{\partial t}, \label{eq:Faraday}\\
  \nabla \times \bm{H} &= \frac{1}{c} \frac{\partial \bm{D}}{\partial t}. \label{eq:Ampere}
\end{align}
This vector formulation was first given by Heaviside in 1885 \cite{rautio2014long}.
Here, we have used the Gaussian unit system.
There are also a tensor formulation of Maxwell's equations,
which is useful when analysing relativistic systems because it is invariant under the Lorentz transformation \cite{landau1971classical}.

\par
The four equations (\ref{eq:Gauss}-\ref{eq:Ampere}) can be reduced to two by introducing Riemann-Silberstein vector \cite{silberstein1907nachtrag, silberstein1907elektromagnetische, silberstein1914theory},
and further reduced to one equation in matrix form by introducing `an optical spinor' \cite{darwin1932notes}.
The form of the matrix-vector equation is the same as that of Dirac equation of a free electron.
That is why it is called `optical Dirac equation' and reveals correspondence between optics and quantum theory.
For instance, there are several works to find the wave function of a photon \cite{bialynicki1994wave, sipe1995photon, bialynicki1996photon, smith2007photon}.

\par
It is pointed out that the photon wave function has nonlocal characteristics \cite{bialynicki1996photon};
nevertheless, there are analogies between classical electromagnetism and quantum mechanics.
Quantum-like calculation can be performed by using this formulation in free space \cite{berry1990quantum, barnett2014optical, bliokh2014extraordinary},
in inhomogeneous anisotropic media,
and in media with the presence of magneto-electric coupling \cite{zalesny2009applications, silveirinha2017p}.
Berry phase effects associated with the optical Dirac Hamiltonian have also been studied \cite{silveirinha2015chern, silveirinha2016z, silveirinha2016bulk, gangaraj2017berry, horsley2018topology}.
In some literatures, dispersion correction has also been taken into account.
To do so, the group permittivity and permeability are used in some works \cite{bliokh2017optical, bliokh2017optical_new_j_phys, bekshaev2018spin},
while the simultaneous equations of the field equation and the equation of electrons motions are considered in other literatures,
where the dissipation in the materials can also be calculated by the perturbation method \cite{raman2010photonic, raman2011perturbation, luan2011electromagnetic, shin2012instantaneous, raman2013upper}.

\par
Starting from the Dirac-like reformulation of classical electromagnetism,
we introduce the concept of `optical density operator', in order to analyse electromagnetic phenomena at finite temperature. 
This paper is organised as following.
In Sec. \ref{sec:density_operator}, beginning with the optical Dirac equation,
we introduce the optical analogue of the density operator in order to take the concept of temperature into account in classical electromagnetism.
We also derive the `optical Liouville-Neumann equation', which is the dynamical equation of the optical density operator.
In Sec. \ref{sec:diagonalisation}, the diagonalisation of the optical Dirac Hamiltonian is shown,
where we decompose the $6\times6$ optical Dirac Hamiltonian into a $2\times2$ matrix and a $3\times3$ matrix.
In Sec. \ref{sec:thermal_state}, we find the thermal state of electromagnetic field and analyse its asymptotic behaviour.
The conclusion is drawn in Sec. \ref{sec:conclusion}.

\section{An optical density operator}
\label{sec:density_operator}

\par
We briefly review how to get the optical Dirac equation from the four Maxwell's equations (\ref{eq:Gauss})--(\ref{eq:Ampere}).
We can regard (\ref{eq:Gauss}) and (\ref{eq:monopole}) as a set of boundary conditions for the electromagnetic field,
while (\ref{eq:Faraday}) and (\ref{eq:Ampere}) are responsible for the dynamics of the field.
By defining `the optical spinor'
\begin{equation}
  \ket{\psi} \equiv \sqrt{\frac{g}{2}}\ 
  \begin{pmatrix}
    \sqrt{\mu}\bm{H} \\ 
    \sqrt{\varepsilon} \bm{E} 
  \end{pmatrix},
  \label{eq:spinor}
\end{equation}
where $g=(4\pi)^{-1}$ is a Gaussian unit coefficient,
we can rewrite the two dynamical equations, (\ref{eq:Faraday}) and (\ref{eq:Ampere}),
in the Dirac-like form,
\begin{align}
  \mathcal{H} \ket{\psi} &= i\frac{\partial}{\partial t} \ket{\psi},
  \mathcal{H} = 
  \begin{pmatrix}
    0 & -i v \nabla \times \\
    i v \nabla \times& 0
  \end{pmatrix}.
  \label{eq:optical_Dirac}
\end{align}

Equation (\ref{eq:optical_Dirac}) is what we call `optical Dirac equation' \cite{barnett2014optical, horsley2018topology} because of the similarity to the Dirac equation of a free electron.
Here,
$v=c/n$ is the speed of light in the medium with a refractive index of $n = \sqrt{\varepsilon \mu}$.
The `optical Dirac Hamiltonian' $\mathcal{H}$ is a Hermitian operator.
Here, we have assumed linear constitutive equations,
\begin{align}
  \begin{cases}{}
    \bm{D} &= \varepsilon \bm{E},\\
    \bm{B} &= \mu \bm{H},
  \end{cases}
\end{align}
where $\varepsilon$ and $\mu$ are the permittivity and the permeability in the medium, respectively.
The optical Dirac equation for a monochromatic field, which has a time harmonic dependence $\exp(-i\omega t)$, is
\begin{align}
  \mathcal{H} \ket{\psi_\omega} = \omega \ket {\psi_\omega}.
\end{align}
We normalise the optical spinor so that
\begin{align}
  \Braket{\psi|\psi} &= \int \mathrm{d} \bm{r}^3\ \frac{g}{2} \left( \varepsilon |\bm{E}|^2 + \mu |\bm{H}|^2 \right) = 1.
\end{align}
This implies that the mean value of the Hamiltonian is the energy of the field with $\hbar = 1$.
Since our optical Dirac Hamiltonian is a Hermitian operator,
the eigenfunctions $\ket{\psi_\omega}$ form an orthonormal basis, $\braket{\psi_{\omega'}|\psi_{\omega}} = \delta(\omega-\omega')$.

\par
We can construct our `optical density operator' with the optical spinor and its Hermitian conjugate.
\begin{equation}
  \rho \equiv \ket{\psi}\bra{\psi} = 
  \frac{g}{2}
  \begin{pmatrix}
    \mu \bm{H}\bm{H}^\dagger & n \bm{H}\bm{E}^\dagger \\
    n \bm{E}\bm{H}^\dagger & \varepsilon \bm{E}\bm{E}^\dagger
  \end{pmatrix}.
  \label{eq:density_operator}
\end{equation}
The diagonal elements of this density operator are the spectral coherence matrices,
which are commonly used to analyse the polarisation or the coherence of light in the nonparaxial regime \cite{setala2002degree, dennis2004geometric, tervo2003degree}.
On the other hand, the offdiagonal elements are the correlation matrices,
which represent the correlation between the electric field and the magnetic field.
Note that the trace of the density operator is normalised to unity,
\begin{align}
  \operatorname{tr} \left( \rho \right) &= \operatorname{tr} \left ( \ket{\psi} \bra{\psi} \right) = 1.
\end{align}

\par
Simply taking the time derivative of the density operator (\ref{eq:density_operator}) and using (\ref{eq:optical_Dirac}),
we can derive the time evolution equation of the density operator,
\begin{equation}
  \frac{\partial \rho}{\partial t} = i\left[\rho, \mathcal{H} \right],
  \label{eq:optical_Liouville}
\end{equation}
which we call the optical Liouville-Neumann equation because of the similarity to the Liouville-Neumann equation in quantum mechanics.
Here, $[\ \circ\ ,\ \bullet\ ]$ is the commutation relation.
This optical Liouville-Neumann equation is a natural extension of Maxwell's equations,
which we can use to calculate partially coherent light.
That is, we can deal not only with coherent superposition but also with incoherent mixture.

\section{Diagonalisation of the optical Dirac Hamiltonian}
\label{sec:diagonalisation}
\par
We can decompose the Hamiltonian,
\begin{align*}
  \mathcal{H} &= \begin{pmatrix}
    0 & -i v \nabla \times \\
    i v \nabla \times & 0
  \end{pmatrix} = \hat{\sigma}_y \otimes v \nabla \times = \hat{\sigma}_y \otimes (-i v\bm{\tau} \cdot \nabla),
\end{align*}
and use the momentum representation,
\begin{align}
  \mathcal{H}_{\bm{k}} &= \hat{\sigma}_y \otimes ( v \bm{\tau} \cdot \bm{k} ).
\end{align}
This Hamiltonian is a matrix which acts on $\mathbb{C}^2 \otimes \mathbb{C}^3$.
Here, we use one of Pauli matrices,
\begin{align*}
  \hat{\sigma}_y &=
  \begin{pmatrix}
    0 & -i\\
    i & 0
  \end{pmatrix},
\end{align*}
and $\bm{\tau}=(\tau_x, \tau_y, \tau_z)^\top$ is a vector whose elements are spin-1 matrices \cite{SchiffLeonardI1968Qm}
\begin{align}
  \tau_x &= 
  \begin{pmatrix}
    0 & 0 & 0\\
    0 & 0 & -i\\
    0 & i & 0
  \end{pmatrix},\quad
  \tau_y = 
  \begin{pmatrix}
    0 & 0 & i\\
    0 & 0 & 0\\
    -i & 0 & 0
  \end{pmatrix},\quad
  \tau_z = 
  \begin{pmatrix}
    0 & -i & 0\\
    i & 0 & 0\\
    0 & 0 & 0
  \end{pmatrix},
\end{align}
in order to represent the vector product in matrix form as in the literatures \cite{berry2009optical, barnett2014optical, bliokh2014extraordinary}.

\par
We consider the following two eigenvalue problems to diagonalise $\mathcal{H}_{\bm{k}}$.
Firstly, for the operator $\hat{\sigma}_y$ acting on $\mathbb{C}^2$,
\begin{align}
\hat{\sigma}_y \bm{d}_{\pm} &= \pm \bm{d}_{\pm},
\end{align}
where the eigenvectors are
\begin{align}
  \bm{d}_{\pm} &= \frac{1}{\sqrt{2}}\begin{pmatrix}
    1\\
    \pm i
  \end{pmatrix}.
\end{align}
These two vectors form a basis of $\mathbb{C}^2$.
Secondly, for the operator $(v\bm{\tau} \cdot \bm{k})$ on $\mathbb{C}^3$, we have
\begin{align}
    (v\bm{\tau} \cdot \bm{k}) \bm{e}_{0}(\bm{k}) &= 0,\\
    (v\bm{\tau} \cdot \bm{k}) \bm{e}_{\pm} (\bm{k}) &= \pm vk \bm{e}_{\pm} (\bm{k}),
\end{align}
where $k = |\bm{k}|$ is the absolute value of the wavevector. 
The corresponding eigenvectors are
\begin{align}
    \bm{e}_{0} (\bm{k}) &= \cfrac{\bm{k}}{k}, \\
    \bm{e}_{\pm} (\bm{k}) &= \cfrac{\bm{k} \times \bm{k} \times \bm{u}_z}{2k^2} \pm i \cfrac{ \bm{k} \times \bm{u}_z}{2k}.
\end{align}
The subscripts, `$0$' and `$\pm$,' label the longitudinal mode and the two transverse modes, respectively.
`$+$, and `$-$' are the left-handed and the right-handed circularly polarised modes.
Note that these three vectors form a basis since they are orthogonal of $\mathbb{C}^2$ to each other because of the hermiticity of $v \bm{\tau} \cdot \bm{k}$.
\par
We can get the six eigenvalues of the Hamiltonian $\mathcal{H}_{\bm{k}}$ by pairing up the eigenvalues of $\sigma_y$ with that of $v \bm{\tau} \cdot \bm{k}$,
and can obtain the six corresponding eigenvectors of $\mathcal{H}_{\bm{k}}$ by combining the corresponding eigenvectors.
The eigenvalue equation of the Hamiltonian is
\begin{align}
  \mathcal{H}_{\bm{k}} \Ket{\psi_j (\bm{k})} = \Omega_j \Ket{\psi_j (\bm{k})},
\end{align}
and the index $j$, the label pair $(\alpha, \beta)$, the eigenvector $\Ket{\psi_j}$, and the eigenvalue $\Omega_j$ are summarised in TABLE \ref{tab:eigensystem}.
\begin{table}[tbp]
  \centering
  \caption{Eigenvalues, eigenvectors of the optical Dirac Hamiltonian $\mathcal{H}_{\bm{k}}$}
  \label{tab:eigensystem}
  \begin{tabular}{c|c|c|c}
    $j$ & $(\alpha, \beta)$ & Eigenvector $\Ket{\psi_j}$ & Eigenvalue $\Omega_j$ \\ \hline \hline
    0 & $(+,0)$ & $\bm{d}_+ \otimes \bm{e}_0$ & 0 \\
    1 & $(-,0)$ & $\bm{d}_- \otimes \bm{e}_0$ & 0 \\
    2 & $(+,+)$ & $\bm{d}_+ \otimes \bm{e}_+$ & $+vk$ \\
    3 & $(-,-)$ & $\bm{d}_- \otimes \bm{e}_-$ & $+vk$ \\
    4 & $(+,-)$ & $\bm{d}_+ \otimes \bm{e}_-$ & $-vk$ \\
    5 & $(-,+)$ & $\bm{d}_- \otimes \bm{e}_+$ & $-vk$
  \end{tabular}
\end{table}
We can use the unitary matrix which have these eigenvectors in the components,
\begin{align}
  \mathcal{P} &=
  \begin{pmatrix}
    \Ket{\psi_0} & \Ket{\psi_1} & \Ket{\psi_2} & \Ket{\psi_3} & \Ket{\psi_4} & \Ket{\psi_5}
  \end{pmatrix},
\end{align}
and its Hermitian conjugate,
\begin{align}
  \mathcal{P}^\dagger &=
  \begin{pmatrix}
    \Bra{\psi_0} \\
    \Bra{\psi_1} \\
    \Bra{\psi_2} \\
    \Bra{\psi_3} \\
    \Bra{\psi_4} \\
    \Bra{\psi_5}
  \end{pmatrix}
\end{align}
for the diagonalisation of the Hamiltonian $\mathcal{H}_{\bm{k}}$.
We can write $\mathcal{H}_{\bm{k}} \mathcal{P} = \mathcal{P} \mathcal{N}$,
and complete the diagonalisation,
\begin{align}
 &\mathcal{P}^\dagger \mathcal{H}_{\bm{k}} \mathcal{P} = \mathcal{N},\\
 &\mathcal{N} = \mathrm{diag}(\Omega_0, \Omega_1, \Omega_2, \Omega_3, \Omega_4, \Omega_5).
\end{align}

\section{Thermal state}
\label{sec:thermal_state}
\par
We can consider so-called the thermal state of an electromagnetic field,
which is a steady state solution to the optical Liouville-Neumann equation (\ref{eq:optical_Liouville}).
The thermal state is defined by
\begin{align}
  \rho_{em}^{th} &= \frac{e^{-\beta \mathcal{H}_{\bm{k}}}}{\operatorname{tr} \left( e^{-\beta \mathcal{H}_{\bm{k}}} \right)}.
\end{align}
Here, $\beta = (k_B T)^{-1}$ is the inverse temperature of the medium where the field lives.
Indeed, we can easily confirm that this is the steady state solution,
since the density matrix of the thermal state and the Hamiltonian are commutable $\left(0 = i\left[\rho_{em}^{th}, \mathcal{H}_{\bm{k}}\right]\right)$.

\par
To get the explicit expression of the thermal state,
we need to evaluate
\begin{align}
  e^{-\beta \mathcal{H}_{\bm{k}}} = \sum_{m=0}^\infty \frac{1}{m!} (-\beta)^m \left(\mathcal{H}_{\bm{k}}\right)^m.
\end{align}
Since we have already diagonalised the Hamiltonian,
we can easily get
\begin{align}
  &\left( \mathcal{H}_{\bm{k}} \right)^m = \mathcal{P}\mathcal{N}^m  \mathcal{P}^\dagger \notag \\
  &= \sum_{j=0}^5 \Ket{\psi_j} (\Omega_j)^m \Bra{\psi_j} = \sum_{j=0}^5 (\Omega_j)^m \Ket{\psi_j}\Bra{\psi_j}.
\end{align}
Therefore, we can evaluate the trace,
\begin{align}
  &\operatorname{tr} \left( e^{-\beta \mathcal{H}_{\bm{k}}} \right)
  = \operatorname{tr} \left[ \sum_{m=0}^\infty \frac{1}{m!} (-\beta)^m \left( \mathcal{H}_{\bm{k}} \right)^m \right] \notag \\
  &= \sum_{m=0}^\infty \frac{1}{m!} (-\beta)^m \sum_{j=2}^5 (\Omega_j)^m \operatorname{tr} \left(\Ket{\psi_j}\Bra{\psi_j}\right) \notag \\
  %% since \Omega_{0,1} = 0, these modes do not contribute to the Hamiltonian, and we can start with j = 2, ignoring j = 0, 1.
  &= \sum_{j=2}^5 \sum_{m=0}^\infty \frac{1}{m!} (-\beta \Omega_j)^m
  = \sum_{j=2}^5 e^{-\beta \Omega_j}\\
  &= e^{-\beta vk} + e^{-\beta vk} + e^{\beta vk} + e^{\beta vk} \notag \\
  &= 4\cosh (\beta vk),
\end{align}
and the numerator,
\begin{align}
  &\sum_{j=0}^5 \sum_{m=0}^\infty \frac{1}{m!} (-\beta)^m (\Omega_j)^m \Ket{\psi_j}\Bra{\psi_j} \notag \\
  &= \left( e^{-\beta vk} \bm{d}_+ \bm{d}_+^\dagger + e^{\beta vk} \bm{d}_- \bm{d}_-^\dagger \right) \otimes \bm{e}_+ \bm{e}_+^\dagger \notag \\
  &\quad + \left( e^{\beta vk} \bm{d}_+ \bm{d}_+^\dagger + e^{-\beta vk} \bm{d}_- \bm{d}_-^\dagger \right) \otimes \bm{e}_- \bm{e}_-^\dagger \notag \\
  &= \frac{1}{2} \left\{
    e^{-\beta vk}
    \begin{pmatrix}
      1 & -i\\
      i & 1
    \end{pmatrix} +
    e^{\beta vk}
    \begin{pmatrix}
      1 & i\\
      -i & 1
    \end{pmatrix}
  \right\} \otimes \bm{e}_+ \bm{e}_+^\dagger \notag \\
  &\quad +
  \frac{1}{2} \left\{
    e^{\beta vk}
    \begin{pmatrix}
      1 & -i\\
      i & 1
    \end{pmatrix} +
    e^{-\beta vk}
    \begin{pmatrix}
      1 & i\\
      -i & 1
    \end{pmatrix}
  \right\} \otimes \bm{e}_- \bm{e}_-^\dagger \notag \\
  &= R(-i\beta vk) \otimes \bm{e}_+ \bm{e}_+^\dagger + R(i\beta vk) \otimes \bm{e}_- \bm{e}_-^\dagger.
\end{align}
Note that $j = 0, 1$ terms do not contribute because the eigenvalues, $\Omega_0$ and $\Omega_1$, are zero.
$R(\theta)$ is the rotation matrix acting on $\mathbb{C}^2$.
\begin{align}
  R(\theta) = \begin{pmatrix}
    \cos \theta & -\sin \theta \\
    \sin \theta & \cos \theta
  \end{pmatrix}.
\end{align}
Finally, we can obtain the representation of the thermal state of an electromagnetic field at the inverse temperature $\beta$
\begin{align}
  \rho_{em}^{th} &= \frac{1}{4\cosh(\beta vk)} \times \notag \\
                 &\quad \left( 
                   R(-i\beta vk) \otimes \bm{e}_+ \bm{e}_+^\dagger
                   + R(i\beta vk) \otimes \bm{e}_- \bm{e}_-^\dagger 
                 \right).
\end{align}

\par
\begin{figure}[htbp]
  \centering
  \includegraphics[width=.5\linewidth]{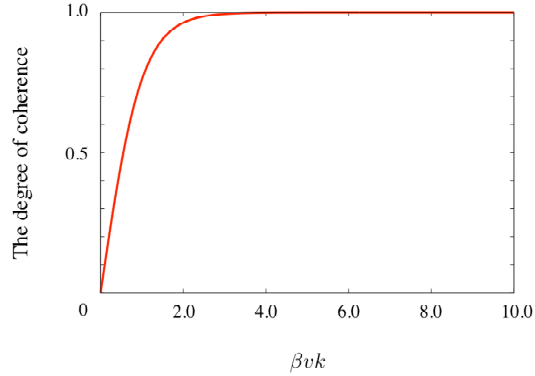}
  \caption{
    Temperature dependence of the correlation (coherence) between the electric field and the magnetic field.
    The coherence is represented by one of the absolute values of the offdiagonal elements of the $2 \times 2$ matrix $R(i\xi)$,
    which consists the optical density operator.
    Note that both of the absolute values of the offdiagonal elements are same.
    At the low temperature limit ($\beta \rightarrow \infty$), the correlation sustains, while vanishes at the high temperature limit ($\beta \rightarrow 0$).
  }
  \label{fig:coherence}
\end{figure}

As we have already seen, we can decompose the density operator into a matrix acting on $\mathbb{C}^2$ and a matrix acting on $\mathbb{C}^3$.
The offdiagonal elements of the matrix on $\mathbb{C}^2$ represent the correlation (coherence) between the electric field and the magnetic field.
All of their absolute values are same.
In FIG. \ref{fig:coherence}, the temperature dependence of the coherence is plotted.
As the temperature rises from the low temperature limit,
the coherence decreases,
and completely vanishes at the high temperature limit.
We check the asymptotics of the density matrix finally.
At the low temperature limit ($\beta \rightarrow \infty$),
we have $\tanh(\beta v k) \rightarrow 1$,
and thus the coherence sustains.
\begin{align}
  \rho_{em}^{th} \rightarrow \frac{1}{4} \begin{pmatrix}
    1 & i\\
    -i & 1
  \end{pmatrix} \otimes \bm{e}_+ \bm{e}_+^\dagger + 
\frac{1}{4} \begin{pmatrix}
    1 & -i\\
    i & 1
  \end{pmatrix} \otimes \bm{e}_- \bm{e}_-^\dagger.
  \label{eq:low_temperature}
\end{align}
Here, we have the coherence factor $i$ in the offdiagonal elements of the $2 \times 2$ matrix in (\ref{eq:low_temperature}).
This factor result from the fact that electromagnetic waves with a wavevector of $\bm{k}$ survive at the low temperature limit.
There is the imaginary unit factor in the relationship between the electric field and the magnetic field of a single circularly-polarised electromagnetic wave ($\bm{H}_{\pm} \propto -i \bm{E}_{\pm}$).
That is where the coherence factor $i$ comes from.
In contrast, at the high temperature limit ($\beta \rightarrow 0$),
the coherence is lost
\begin{align}
  \rho_{em}^{th} &\rightarrow \frac{1}{4} \identity_2 \otimes \bm{e}_+ \bm{e}_+^\dagger 
  + \frac{1}{4} \identity_2 \otimes \bm{e}_- \bm{e}_-^\dagger,
\end{align}
where $\identity_2$ is the identity matrix acting on $\mathbb{C}^2$.
Contrary to the low temperature limit,
the offdiagonal elements of the $2 \times 2$ matrix are zero.
This means that there is no correlation between the electric field and the magnetic field,
and thus no electromagnetic waves can survive but the electric field and the magnetic field exist independently.
In other words, the electric field and the magnetic field are decoupled from each other.
This is the loss of coherence of electromagnetic field at the high temperature limit.

\par
The characteristic scale of the temperature at which the decoherence become significant is determined from $\beta vk$.
The correlation between electric field and magnetic field gradually decreases as $\beta vk$ become small,
and sharply drops around $\beta vk=1$,
where $\beta=(k_B T)^{-1}$ is the inverse temperature,
and $vk$ is the energy of the field in the unit $\hbar=1$.
Thus, the decoherence become significant when the temperature of the environment is comparable to the energy of the field,
e.g. $1.16045 \times 10^4\ \mathrm{K}$ for visible light ($\approx 1 \mathrm{eV}$).
The nature of the decoherence is thermal disturbance by the environment.
As mentioned above, the decoherence occurs at the energy scale where the field energy and that of the environment are comparable.
The decoherence of two fields is analogous to the decoherence process of a two-level quantum system which is resonantly coupled to a thermal bath \cite{rivas2012open}.

\section{Conclusion}
\label{sec:conclusion}
\par
In order to approach electromagnetic phenomena at finite temperature,
we introduced the optical density operator which is constructed of the optical spinor (the wave function of an electromagnetic field) in this paper.
We revealed how the electromagnetic vector field behaves,
focusing on the coherence of the field in the thermal equilibrium at finite temperature.

\par
Starting from Dirac-equation-like reformulation of electromagnetism,
we derived the evolution equation of the density operator.
Since the evolution equation has the same form as the Liouville-Neumann equation in quantum mechanics,
we call it `the optical Liouville-Neumann equation.'
We also found the existence of the thermal state which is a steady state solution to the optical Liouville-Neumann equation.

\par
The thermal state is the incoherent mixture of the left-handed mode and the right-handed mode with the equal weight.
According to the asymptotic analysis of the thermal state,
we found that the electromagnetic field could keep its coherence at the low temperature limit ($\beta \rightarrow \infty$),
while the field lost its coherence at high temperature limit ($\beta \rightarrow 0$).
This means that there is no longer electromagnetic waves but independent electric fields and magnetic fields at the high temperature limit.

\par
This work opens the door to a possibility of the analysis of electromagnetic phenomena at finite temperature in the framework of classical electromagnetism.

\section*{Acknowledgement(s)}
I thank Yiming Lai and Taiki Matsushita for fruitful discussions.
I also thank Samuel Palmer for editing the English text of a draft of this manuscript.
Here, I specially celebrate the 60$^{\mathrm{th}}$ birthday of Prof. Hajime Ishihara who always kindly supports me.

\section*{Funding}
D.O. is funded by the President's PhD Scholarships at Imperial College London.

%\bibliographystyle{tfp}
%\bibliography{manuscript}

\end{document}